\documentclass[aps,prl,twocolumn,showpacs,amsmath,amsmath,amssymb]{revtex4-1}

\usepackage{amsmath,amssymb,graphicx,bm,psfrag,bbold,float,color}

\usepackage[mathscr]{eucal}

\DeclareMathAlphabet{\mathpzc}{OT1}{pzc}{m}{it}

\begin{document}

\title{Superconducting Resonator-Rydberg Atom Hybrid in the Strong Coupling Regime}

\author{Deshui Yu$^{1}$, Alessandro Landra$^{1}$, Mar\'ia Mart\'inez Valado$^{1}$, Christoph Hufnagel$^{1}$, Leong Chuan Kwek$^{1,2,3,4}$, Luigi Amico$^{1,5,6}$, and Rainer Dumke$^{1,7}$}

\email{rdumke@ntu.edu.sg}

\affiliation{$^{1}$Centre for Quantum Technologies, National University of Singapore, 3 Science Drive 2, Singapore 117543, Singapore}

\affiliation{$^{2}$Institute of Advanced Studies, Nanyang Technological University, 60 Nanyang View, Singapore 639673, Singapore}

\affiliation{$^{3}$National Institute of Education, Nanyang Technological University, 1 Nanyang Walk, Singapore 637616, Singapore}

\affiliation{$^{4}$MajuLab, CNRS-UNS-NUS-NTU International Joint Research Unit, UMI 3654, Singapore}


\affiliation{$^{5}$CNR-MATIS-IMM \& Dipartimento di Fisica e Astronomia, Universit\'a Catania, Via S. Soa 64, 95127 Catania, Italy}

\affiliation{$^{6}$INFN Laboratori Nazionali del Sud, Via Santa Sofia 62, I-95123 Catania, Italy}

\affiliation{$^{7}$Division of Physics and Applied Physics, Nanyang Technological University, 21 Nanyang Link, Singapore 637371, Singapore}

\begin{abstract}
We propose a promising hybrid quantum system, where a highly-excited atom strongly interacts with a superconducting LC oscillator via the electric field of capacitor. An external electrostatic field is applied to tune the energy spectrum of atom. The atomic qubit is implemented by two eigenstates near an avoided-level crossing in the DC Stark map of Rydberg atom. Varying the electrostatic field brings the atomic-qubit transition on- or off-resonance to the microwave resonator, leading to a strong atom-resonator coupling with an extremely large cooperativity. Like the nonlinearity induced by Josephson junctions in superconducting circuits, the large atom-resonator interface disturbs the harmonic potential of resonator, resulting in an artificial two-level particle. Different universal two-qubit logic gates can also be performed on our hybrid system within the space where an atomic qubit couples to a single photon with an interaction strength much larger than any relaxation rates, opening the door to the cavity-mediated state transmission.
\end{abstract}


\maketitle

\textit{Introduction.} Superconducting (SC) circuits and neutral atoms define coherent systems of key importance in quantum technology. Besides the features of flexibility, tunability, and scalability, SC devices manipulate quantum states rapidly owing to the strong coupling to external fields. Nevertheless, this sensitivity of SC circuits leads to short decoherence times caused by environmental noise~\cite{PRL:Stern2014}. In contrast, atoms can maintain the quantum coherence exceeding one second~\cite{PRL:Treutlein2004}, but processing quantum information as fast as SC devices is impractical.

Hybridizing SC circuits and atoms bears great potential to overcome the bottlenecks of above quantum technologies~\cite{Nature:Colombe2007,PRA:Siercke2012,Nature:Bernon,RMP:Xiang2013,PRA:Yu2016}. In this context, one competitive candidate is the SC resonator-atom system, where a coherent microwave photon strongly drives an atomic transition between two long-lived states~\cite{PRA:Petrosyan2009,PRA:Sarkany2015}. 
The SC resonators, such as coplanar waveguide (CPW) cavity and LC resonator, can maintain the quantum coherence of microwave photons of the order of 1 $\mu$s - 1 ms~\cite{PRA:Blais2004}. The strong hybrid coupling requires the microscopic particles possess large dipole moments and long-lifetime qubit states, for which atoms in highly-excited Rydberg states are usually employed as intermediate qubits to interact with the resonator~\cite{RMP:Saffman2010}. After gate operations, the quantum information encoded in Rydberg states can be mapped onto two hyperfine ground states for long-time storage~\cite{PRA:Pritchard2014}. Moreover, the energy spectrum of highly-excited states can be controlled by the external electrostatic field, making the hybrid system more tunable. 

\begin{figure}
\includegraphics[width=8.5cm]{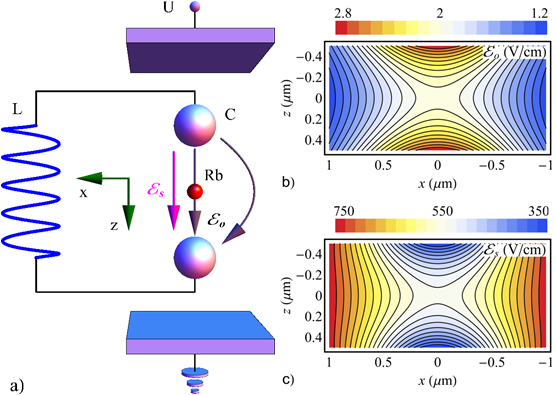}\\
\caption{(Color online) (a) A $^{87}$Rb atom interacts with a LC oscillator. The capacitor $C$ is composed of a pair of equal-sized conducting spheres with a radius of 0.3 $\mu$m and an intercenter distance of 3.1 $\mu$m along the $z$-axis, resulting in $C=20$ aF. The inductance is chosen to be $L=24.8$ $\mu$H, leading to the oscillation frequency $\omega_{0}=2\pi\times7.1$ GHz. The $^{87}$Rb atom is placed at the halfway point (the origin of coordinate) between two spheres and couples to the quantized electric field (the amplitude ${\cal E}_{o}$) of capacitor. A pair of parallel plates (in the $x-y$ plane) with an imposed voltage difference of $U$ is applied to generate an electrostatic-field bias ${\cal E}_{s}$ at the position of atom, inducing the DC Stark shifts of atomic states. Distributions of ${\cal E}_{o}$ and an example of ${\cal E}_{s}$ around the position of atom in the $x-z$ plane are shown in (b) and (c), respectively.}\label{Fig1}
\end{figure}

In analogy with the SC qubits~\cite{Nature:Clarke} relying on Josephson junctions (JJ), the anharmonicity coming from the atom-resonator interface distorts the harmonic oscillators inside cavity, resulting in the dressed-state qubit~\cite{Nature:Timoney2011}. Currently, the decoherence times of SC qubits are strongly limited by the $1/f$ charge noise in JJs, although some have been improved to tens of $\mu$s~\cite{PRL:Paik2011,Science:Wang2016}. Replacing JJs by atoms, as proposed in this work, would be an option to maintain the coherence of SC circuits for a long time.

The Fabry-P\'erot (FP) resonator-Rydberg atom system has been widely employed to explore the fundamentals of quantum optics~\cite{PRL:Rempe1987,PRL:Brune1990,PRL:Brune1996,Nature:Nogues1999,PRL:Rauschenbeutel1999,RMP:Raimond2001,RepProgPhys:Walther2006,Nature:Gleyzes2007,Nature:Guerlin2007}. Recently, relevant research has been extended to the SC artificial atoms strongly coupling to a SC resonator, i.e., circuit QED~\cite{Nature:Wallraff2004,Nature:You2011}. In comparison with the common FP cavity, the extremely small cavity-mode volume of SC resonator results in a large internal electric vacuum field and may lead to a strong SC resonator-Rydberg atom interaction.

In this paper, we propose a promising hybrid system, where a highly-excited Rydberg atom strongly interacts with a SC microwave LC resonator. Referring to an avoided-level crossing in the DC Stark spectroscopy of Rydberg atom, the atomic-qubit transition can be tuned on- or off-resonance to the LC resonator by using an external electrostatic field. As we shall see, the resonant atom-cavity interaction breaks the harmonicity of LC oscillator, leading to an artificial two-level particle without JJs. Moreover, different universal two-qubit logic gates can be implemented on this hybrid system, where an atomic qubit couples to a single photon with an interaction strength much larger than any relaxation rates and an extremely large cooperativity, showing the potential of the cavity-mediated state transmission

\textit{Physical model.} We consider a SC microwave resonator, where an inductor $L=24.8$ $\mu$H connects to a capacitor $C=20$ aF consisting of two identical conducting spheres with a radius of 0.3 $\mu$m and a center-to-center distance of 3.1 $\mu$m in the $z$-direction [Fig.~\ref{Fig1}(a)]. The Hamiltonian of LC resonator is given by $H_{LC}=\hbar\omega_{0}(a^{\dag}a+1/2)$ with the oscillation frequency $\omega_{0}=1/\sqrt{LC}=2\pi\times7.1$ GHz. $a^{\dag}$ and $a$ are the raising and lowering operators with the commutation relation $[a,a^{\dag}]=1$. According to the capacitor charge $Q=i\sqrt{C\hbar\omega_{0}/2}(a^{\dag}-a)$~\cite{PRX:Todorov2014}, the quantized electric field of capacitor has the form ${\bf E}_{o}={\bf e}_{o}i{\cal{E}}_{o}(a^{\dag}-a)/2$, where the unit vector ${\bf e}_{o}$ represents the field direction and the amplitude ${\cal{E}}_{o}$ can be numerically derived using Coulomb's law.

A highly-excited $^{87}$Rb atom placed at the midpoint of capacitor couples to the LC resonator via the electric field ${\bf E}_{o}$ of capacitor [Fig.~\ref{Fig1}(a)]. In the dipole approximation, the atom-resonator interaction is given by $V_{o}=-{\bf D}\cdot{\bf E}_{o}$, where ${\bf D}$ is the atomic dipole moment vector. Additionally, an external electrostatic field ${\bf E}_{s}={\bf e}_{s}{\cal E}_{s}$, generated by an extra parallel-plate (in the $x-y$ plane) capacitor with an imposed voltage difference $U$, is applied to tune the energy spectrum of Rydberg atom. The corresponding atom-field coupling is written as $V_{s}=-{\bf D}\cdot{\bf E}_{s}$.

The numerical result of the amplitude ${\cal{E}}_{o}$ is shown in Fig.~\ref{Fig1}(b). At the position of atom, ${\bf e}_{o}$ is along the $z$-axis and ${\cal{E}}_{o}=2.0$ V/cm. Figure~\ref{Fig1}(c), as an example, displays the distribution of the electrostatic field generated by the parallel-plate capacitor with ${\bf e}_{s}$ along $z$-direction and ${\cal E}_{s}=550.7$ V/cm at the atomic position. For a typical Rydberg-state radius of about 40 nm in this paper, the inhomogeneities of ${\cal{E}}_{o}$ and ${\cal{E}}_{s}$ and the anisotropies of field directions ${\bf e}_{o}$ and ${\bf e}_{s}$ caused by the finite size of atom are all less than 0.2 \% and neglectable. As we will see below, varying ${\cal{E}}_{s}$ (via tuning the voltage difference $U$) brings a pair of atomic states on- or off-resonance to the LC resonator. Finally, we assume the hybrid system operates at the low temperature of $T=20$ mK with the thermal fluctuation of $\omega_{T}=2\pi\times0.4$ GHz.

\begin{figure}[b]
\includegraphics[width=8.5cm]{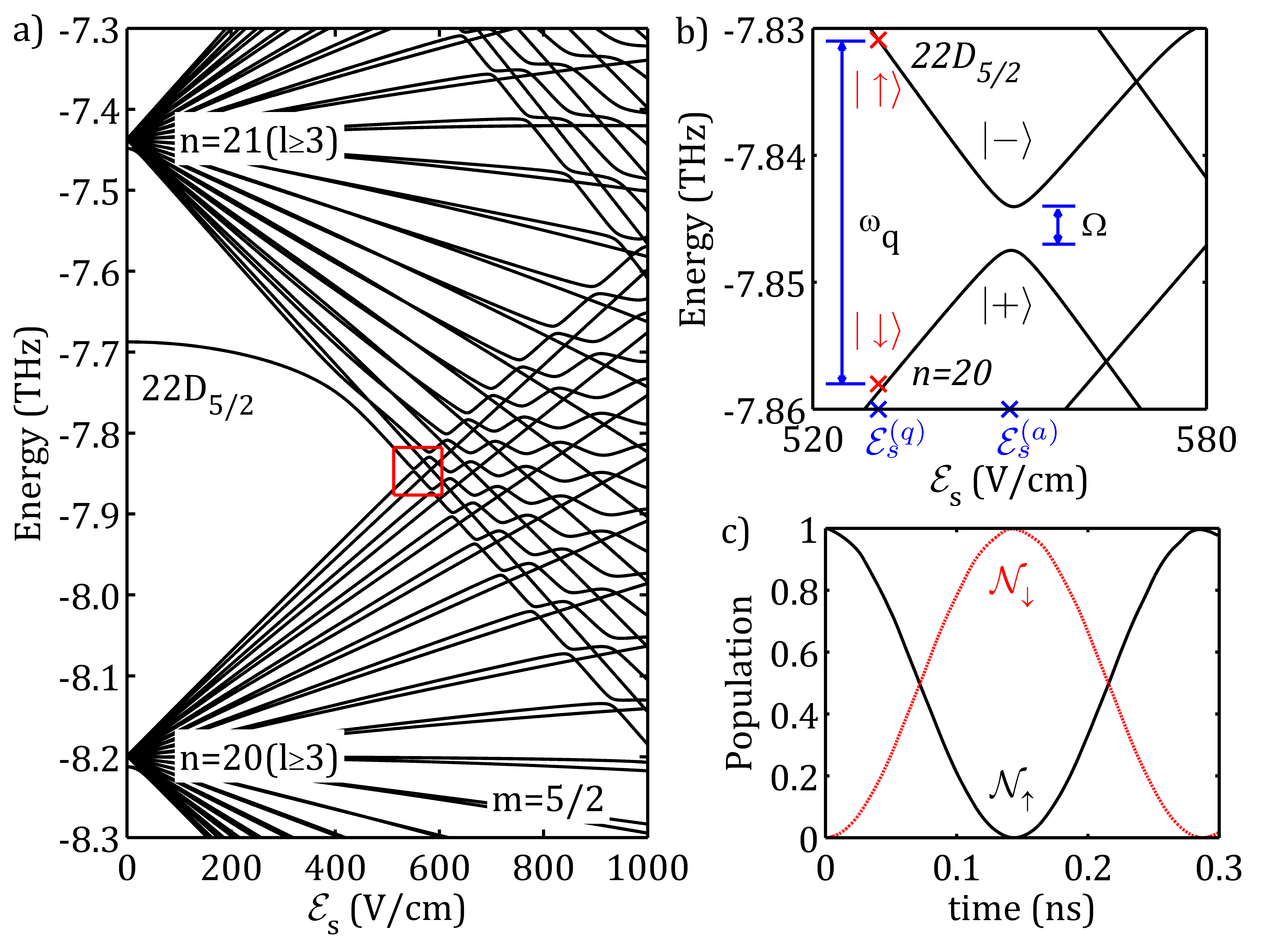}\\
\caption{(Color online) Atomic qubit. (a) DC Stark map in the electrostatic field ${\cal{E}}_{s}$ with the $z$-component of total angular momentum $m=\frac{5}{2}$. The detailed energy spectrum surrounded by the rectangle frame is shown in (b), where an avoided crossing with an energy separation $\Omega=2\pi\times3.2$ GHz occurs at ${\cal E}^{(a)}_{s}\equiv550.7$ V/cm. Two eigenstates $|\downarrow\rangle$ and $|\uparrow\rangle$ at ${\cal E}^{(q)}_{s}\equiv530$ V/cm near the anticrossing are chosen to form an atomic qubit. The qubit-transition frequency is $\omega_{q}=2\pi\times27.1$ GHz. When ${\cal E}_{s}$ is adiabatically reduced to zero, $|\uparrow\rangle$ and $|\downarrow\rangle$ approach $22D_{5/2}$ and a superposition state $\Phi_{n=20}$ combined by a set of $|n=20,l\geq3,j,m=\frac{5}{2}\rangle$, respectively. (c) Vacuum-Rabi oscillation of atom between $|\downarrow\rangle$ and $|\uparrow\rangle$ with ${\cal E}_{s}={\cal E}^{(a)}_{s}$ and the atom initially in $|\uparrow\rangle$. The Rabi frequency is $\Omega$.}\label{Fig2}
\end{figure}

\textit{Atomic qubit.} We first consider the eigenenergies and eigenstates of $^{87}$Rb under the control of ${\bf E}_{s}$ in the absence of ${\bf E}_{o}$ via diagonalizing the Hamiltonian $H=H_{a}+V_{s}$ in the basis of $|nl_{j}(m)\rangle$, where $H_{a}$ represents the free-atom Hamiltonian and $n$, $l$, $j$, and $m$ are the principal, orbital, total angular momentum, and magnetic quantum numbers, respectively. Since ${\cal E}_{s}$ along the $z$-direction keeps $m$ unchanged, we restrict ourselves within the basis of $m=\frac{5}{2}$ as an example. Like the SC qubits~\cite{PRA:Blais2004}, a pair of eigenstates near an avoided-level crossing and far away from other states in energy are employed as two qubit states to encode the quantum information. The qubit transition occurs when ${\cal E}_{s}$ is set at this anticrossing.

Figure~\ref{Fig2}(a) shows the DC Stark map of $^{87}$Rb around $22D_{5/2}(m=\frac{5}{2})$ as a function of ${\cal E}_{s}$, derived according to the methods in~\cite{PRA:Zimmerman1979,PRA:Marinescu1994}. It is seen that there exist plenty of avoided-level crossings. Nonetheless, any anticrossings with the energy spacings smaller than $\omega_{T}$ cannot be observed. We focus on the avoided crossing (labeled by the superscript $a$) at ${\cal E}_{s}={\cal E}^{(a)}_{s}\equiv550.7$ V/cm occurring between two energy curves starting with $22D_{5/2}$ and a superposition state $\Phi_{n=20}$ composed of a set of $|n=20,l\geq3,j,m=\frac{5}{2}\rangle$ at ${\cal E}_{s}=0$ [Fig.~\ref{Fig2}(b)]. The energy separation is as high as $\Omega=2\pi\times3.2$ GHz. The eigenstates $|\downarrow\rangle$ and $|\uparrow\rangle$ of these two energy curves at ${\cal E}_{s}={\cal E}^{(q)}_{s}\equiv530$ V/cm (the superscript $q$ denotes the qubit) near to the anticrossing are far away from any other states and can be chosen to implement an atomic qubit with the transition frequency $\omega_{q}=2\pi\times27.1$ GHz. The population ${\cal N}_{\uparrow}=\langle\sigma_{+}\sigma_{-}\rangle$ ($\sigma_{-}=|\downarrow\rangle\langle\uparrow|$ and $\sigma_{+}=\sigma^{\dag}_{-}$) of the atom in $|\uparrow\rangle$ can be derived from adiabatically reducing ${\cal E}_{s}$ to a low bias ${\cal E}^{(b)}_{s}$ (for example, ${\cal E}^{(b)}_{s}=100$ V/cm) and measuring the atom in $22D_{5/2}(m=\frac{5}{2})$ via the standard spectroscopic techniques. The preparation of the atom in $|\uparrow\rangle$ is implemented via adiabatically increasing ${\cal E}_{s}$ from ${\cal E}^{(b)}_{s}$ to ${\cal E}^{(q)}_{s}$ with the atom initially in $22D_{5/2}$ at ${\cal E}_{s}={\cal E}^{(b)}_{s}$. The radii of atom in $|\downarrow\rangle$, $|\uparrow\rangle$, $22D_{5/2}(m=\frac{5}{2})$, and $\Phi_{n=20}$ are less than 40 nm~\cite{JPB:Low2012}, and the inhomogeneity of ${\cal E}_{s}$ can be neglected.

The qubit transition of atom is performed by nonadiabatically increasing the electrostatic field ${\cal E}_{s}$ from ${\cal E}^{(q)}_{s}$ to ${\cal E}^{(a)}_{s}$. Figure~\ref{Fig2}(c) displays the Rabi oscillations of populations ${\cal N}_{u}$ ($u=\downarrow,\uparrow$ and ${\cal N}_{\downarrow}=\langle\sigma_{-}\sigma_{+}\rangle$) of the atom in $|u\rangle$ with ${\cal E}_{s}={\cal E}^{(a)}_{s}$. It is seen that the total population is almost unity, ${\cal N}_{\downarrow}+{\cal N}_{\uparrow}\approx1$, indicating that the atom is hardly transferred to any other eigenstates and can be effectively simplified as a closed two-state system. The Rabi frequency is equal to the anticrossing gap $\Omega\simeq|\langle\downarrow|V_{s}/\hbar|\uparrow\rangle|$ and the $\pi$-pulse time duration for the state flipping of atom is given by $\pi/\Omega$. Two eigenstates at the avoided crossing are $|\pm\rangle=(|\downarrow\rangle\pm|\uparrow\rangle)/\sqrt{2}$

Within the space spanned by $|\downarrow\rangle$ and $|\uparrow\rangle$ and ${\cal E}^{(q)}_{s}\leq{\cal E}_{s}\leq{\cal E}^{(a)}_{s}$, the system Hamiltonian in the absence of ${\bf E}_{o}$ is reduced as $H/\hbar=\frac{\omega_{q}-\Delta\omega_{DC}}{2}\sigma_{z}+\frac{\Omega}{2}\sigma_{x}$, where the $z$- and $x$-component Pauli matrices $\sigma_{z}=\sigma_{+}\sigma_{-}-\sigma_{-}\sigma_{+}$ and $\sigma_{x}=\sigma_{+}+\sigma_{-}$. $\Delta\omega_{DC}$, which depends on ${\cal E}_{s}$, denotes the DC Stark-shift difference between $|\downarrow\rangle$ and $|\uparrow\rangle$. When ${\cal E}_{s}$ is increased from ${\cal E}^{(q)}_{s}$, $\Delta\omega_{DC}$ increases as well. At ${\cal E}_{s}={\cal E}^{(a)}_{s}$, we have $\omega_{q}=\Delta\omega_{DC}$ and are left with a resonantly-driven two-state system.

\begin{figure}
\includegraphics[width=8.5cm]{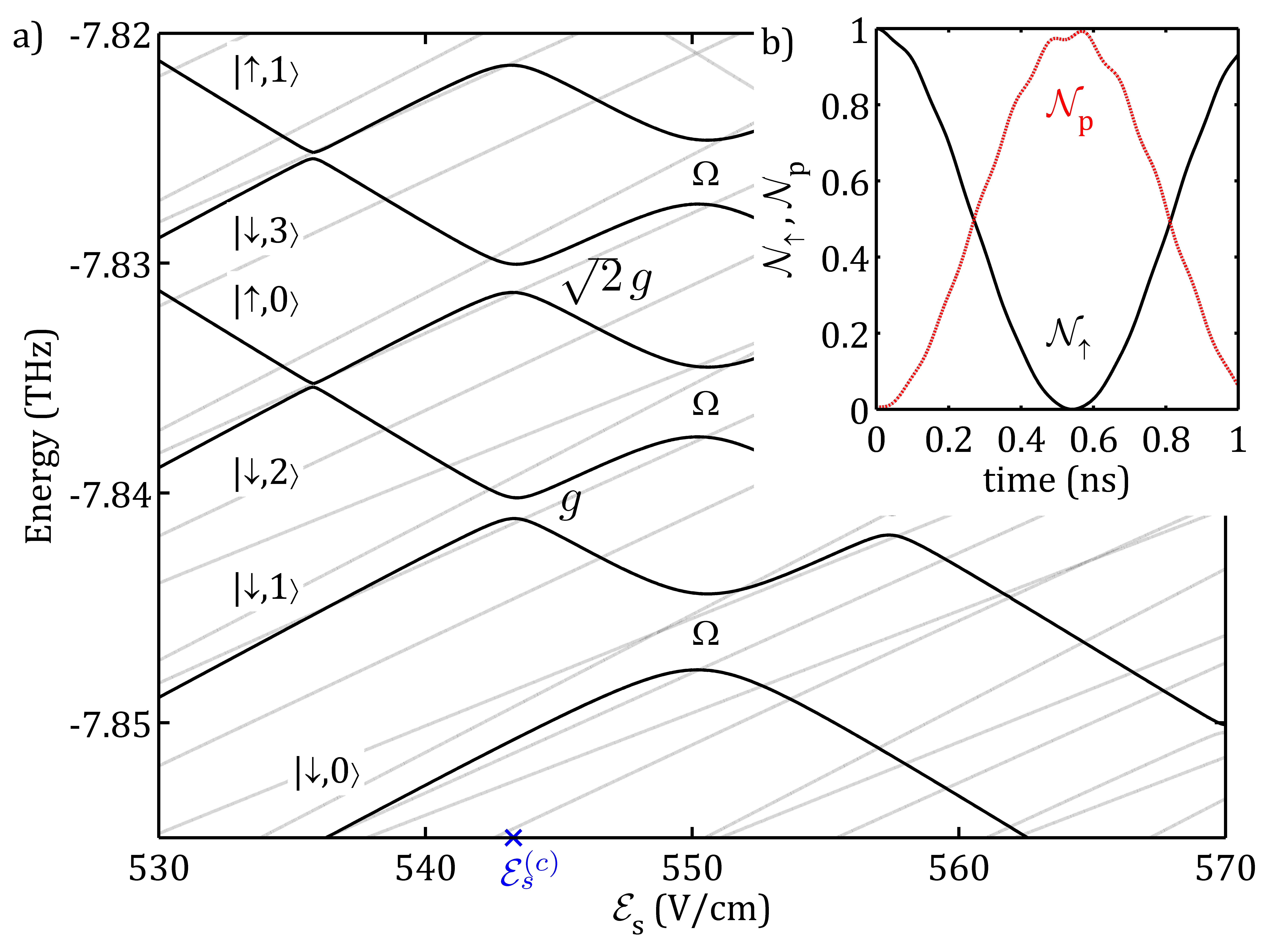}\\
\caption{(Color online) (a) Energy spectrum of hybrid system as a function of the electrostatic field ${\cal{E}}_{s}$. The energy curves associated with $|\downarrow\rangle$ and $|\uparrow\rangle$ are thickened. The energy spacings of avoided-level crossings occurring between $|\downarrow,N+1\rangle$ and $|\uparrow,N\rangle$ at ${\cal E}_{s}={\cal E}^{(c)}_{s}\equiv543.3$ V/cm are $\sqrt{N+1}g$ with $g=2\pi\times0.90$ GHz. (b) Vacuum-Rabi oscillations of ${\cal N}_{\uparrow}$ and ${\cal N}_{p}=\langle a^{\dag}a\rangle$ via solving the time-dependent Schr\"odinger equation with ${\cal E}_{s}={\cal E}^{(c)}_{s}$. The atom is initially prepared in $|\uparrow\rangle$ while no microwave photons exist inside the cavity.}\label{Fig3}
\end{figure}

\textit{Strong atom-resonator coupling.} In the presence of ${\bf E}_{o}$, the atom is dressed by the microwave electromagnetic field. Diagonalizing the system Hamiltonian $H=H_{LC}+H_{a}+V_{s}+V_{o}$ in the basis of $|nl_{j}(m),N\rangle\equiv|nl_{j}(m)\rangle\otimes|N\rangle$, where the integer $N=0,1,2,...$ denotes the number of microwave photons inside the LC resonator, results in the energy spectrum of hybrid system shown in Fig.~\ref{Fig3}(a). As is illustrated, the free Hamiltonian $H_{LC}$ shifts the DC Stark map of Rydberg atom by steps $N\hbar\omega_{o}$. The extra atom-resonator interface $V_{o}$ leads to more avoided-level crossings occurring at the intersections between energy curves with different $N$. Nevertheless, the anticrossings induced by multiphoton processes exhibit the energy separations smaller than $\omega_{T}$ and, hence, can be ignored.

We focus on the eigenenergies associated with the atomic-qubit states $|\downarrow\rangle$ and $|\uparrow\rangle$. As shown in Fig.~\ref{Fig3}(a), the energy spacing of one-photon anticrossing happening between $|\downarrow,N+1\rangle$ and $|\uparrow,N\rangle$ at ${\cal E}_{s}={\cal E}^{(c)}_{s}\equiv543.3$ V/cm (the superscript $c$ denotes the atom-resonator coupling) approximates $\sqrt{N+1}g$, where $g\simeq2|\langle\downarrow,1|V_{o}/\hbar|\uparrow,0\rangle|$ is the single photon-atom interaction strength. For our physical specification, $g$ approximates $2\pi\times0.90$ GHz, larger than any coupling strengths achieved in recently experimentally demonstrated atom-cavity and SC qubit-resonator systems~\cite{PRL:Imamog1999,Nature:Reithmaier2004,PRB:Greuter2015,PRL:Fedorov2010,NewJPhys:Yamamoto2014,NatPhys:Niemczyk2010,PRB:Inomata2012,PRB:Orgiazzi2016,Science:Hood2000,Nature:McKeever2003,Nature:Birnbaum2005,Nature:Aoki2006} and almost equal to that of the atom-waveguide system~\cite{Science:Thompson2013,Nature:Tiecke2014}. Similar to the nonlinearity introduced by integrating JJs into the LC circuit, the strong atom-resonator coupling disturbs the harmonic potential of LC resonator and makes $|\downarrow,N+1\rangle$ and $|\uparrow,N\rangle$ well separated from others around ${\cal E}^{(c)}_{s}$, resulting in the SC dressed-state qubits~\cite{Nature:Timoney2011}. For the pair of $|\downarrow,1\rangle$ and $|\uparrow,0\rangle$, the corresponding eigenstates of hybrid system at ${\cal E}^{(c)}_{s}$ are the maximally entangled states (two Bell states) $|\Psi_{\pm}\rangle=\frac{1}{\sqrt{2}}(|\downarrow,1\rangle\pm|\uparrow,0\rangle)$.

Figure~\ref{Fig3}(b) displays the expectation values of the number ${\cal N}_{p}=\langle a^{\dag}a\rangle$ of photons inside the resonator and the population ${\cal N}_{\uparrow}$ of atom in $|\uparrow\rangle$ as a function of time with the atom initially prepared in $|\uparrow\rangle$ and none photons in resonator at ${\cal E}_{s}={\cal E}^{(c)}_{s}$. It is seen that an energy quantum is being exchanged back and forth between atom and resonator with a vacuum-Rabi frequency of $g$. Moreover, ${\cal N}_{\uparrow}+{\cal N}_{p}\approx1$ indicates that the hybrid system can be simplified as a two-state system composed of $|\downarrow,1\rangle$ and $|\uparrow,0\rangle$, i.e., the dressed-state qubit. The $\pi$-pulse time duration for flipping states of atom and photon is $\pi/g$.

Within the Hilbert space spanned by $\{|u,N\rangle,u=\downarrow,\uparrow;N=0,1,2,...\}$ and ${\cal E}^{(q)}_{s}\leq{\cal E}_{s}\leq{\cal E}^{(a)}_{s}$, the hybrid-system Hamiltonian can be simplified as $H/\hbar=\frac{\omega_{q}-\Delta\omega_{DC}}{2}\sigma_{z}+\omega_{0}a^{\dag}a+\frac{\Omega}{2}\sigma_{x}-i\frac{g}{2}(a^{\dag}-a)\sigma_{x}$. Varying ${\cal E}_{s}$ tunes both the atomic-qubit transition and the atom-resonator interaction. The resonant Rabi oscillation of atom occurs when $\Delta\omega_{DC}=\omega_{q}$. The atom strongly couples to the microwave resonator when $\Delta\omega_{DC}=\omega_{q}-\omega_{0}$.

Typically, the quality factor for the SC microwave LC resonator is of the order of $10^{4}$~\cite{PRA:Blais2004}. For our specification, the loss rate of LC resonator is estimated to be $\kappa=2\pi\times0.7$ MHz, much smaller than $\Omega$ and $g$. In addition, the decoherence rate of the atomic-qubit transition approximates $\gamma=2\pi\times0.16$ MHz when the Rydberg atom is positioned near a metallic surface~\cite{PRA:Crosse2010}, resulting in a hybrid system in the strong-coupling regime and an extremely large cooperativity ${\cal C}=g^{2}/(\kappa\gamma)=7.2\times10^{6}$ (${\cal C}^{-1}$ measures the critical atomic number for lasing dynamics~\cite{Nature:McKeever2003}). Increasing the amplitude ${\cal{E}}_{o}$ can further enhance the vacuum-Rabi frequency $g$, leading to the ultrastrong atom-resonator interface~\cite{PRL:Diaz2010,PRX:Xie}.

We should note that the adsorbates deposited on chip surface can give rise to large electric field as pointed out in~\cite{PRA:Hattermann2012}. However, experimentally, there might be ways to minimize the detrimental effect of stray electric fields. It was shown that the direction of electric field produced by adsorbates due to the chemisorption or physisorption depends on the material properties~\cite{PRL:Chan2014}. In principle, one can envision to pattern the surface with two materials which give rise to opposing dipole moments of adsorbates. Furthermore, as demonstrated in~\cite{PRA:HermannAvigliano2014}, the electric field due to adsorbates can be minimized by saturating the adsorbate film. The remaining uniform electric fields could be canceled by applied offset field. However, there might still be residual field gradients which are hard to predict. Parasitic static electric field gradients could affect the motion of atom confined inside the trapping region. It is hard to give absolute numbers since the electric field gradient due to adsorbates depends on the realized atomic chip and materials being used for the substrate or the superconductor. 

\begin{figure}
\includegraphics[width=8.5cm]{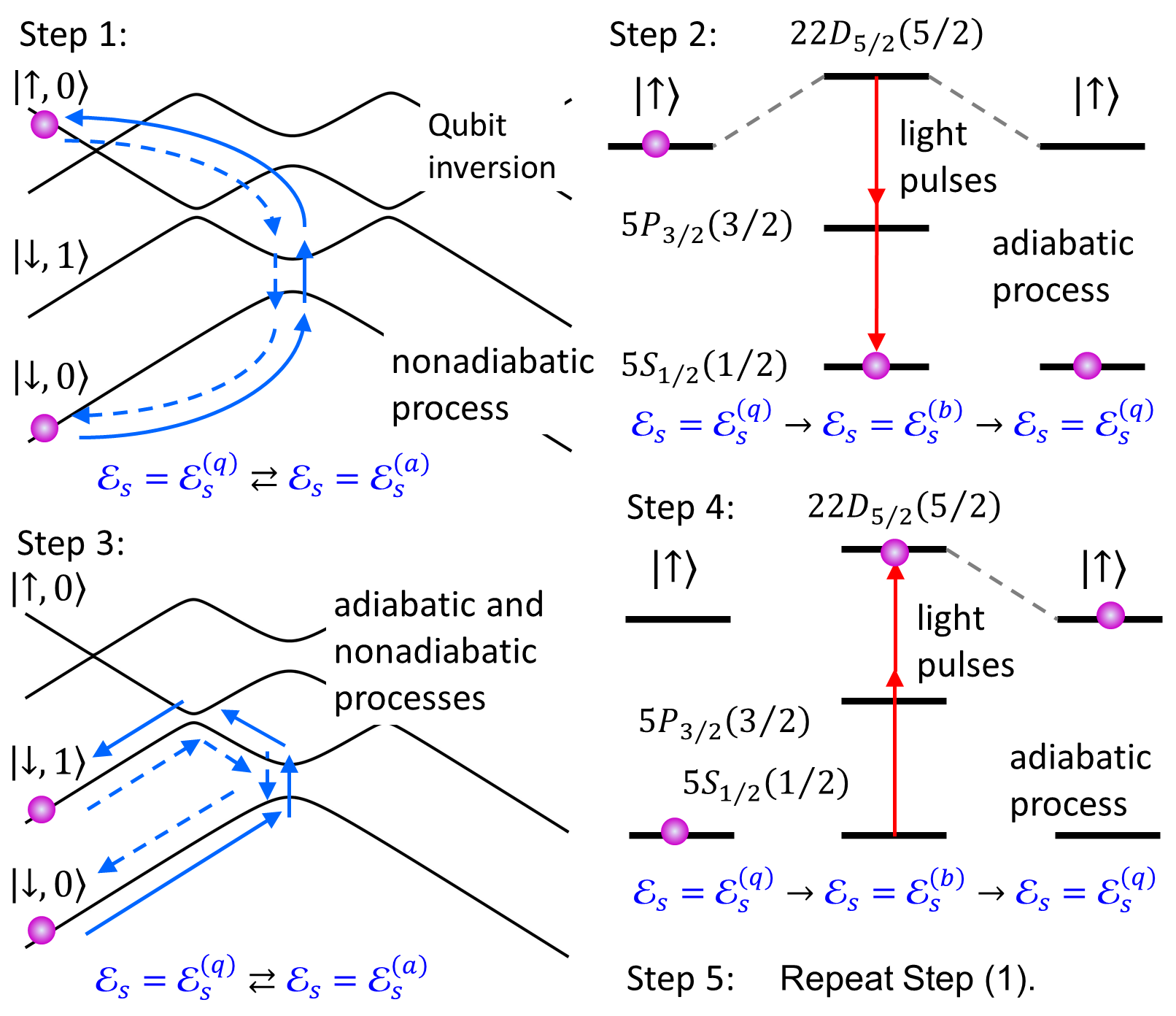}\\
\caption{(Color online) CNOT operation, where the atom acts as the control qubit while the photonic qubit plays the target role. The different steps are discussed in text.}\label{Fig4}
\end{figure}

\textit{Two-qubit logic gate.} Our hybrid system is applicable for two-qubit universal logic gates. We restrict ourselves to the Hilbert space spanned by $|\downarrow,0\rangle$, $|\downarrow,1\rangle$, $|\uparrow,0\rangle$, and $|\uparrow,1\rangle$, i.e., an atomic qubit dressed by a single-photon qubit. As an example, we focus on the controlled-NOT (CNOT) operation, where the photon-state flipping is conditioned on the atomic-qubit state. The concrete implementation (see Fig.~\ref{Fig4}) can be accomplished via the following five steps: (1) The electrostatic field ${\cal E}_{s}$, which is initially set at ${\cal E}^{(q)}_{s}$, nonadiabatically raises to ${\cal E}^{(a)}_{s}$. After staying at ${\cal E}^{(a)}_{s}$ for a $\pi$-pulse time duration of $\pi/\Omega$, ${\cal E}_{s}$ is reduced back to ${\cal E}^{(q)}_{s}$ nonadiabatically. (2) ${\cal E}_{s}$ decreases to the low bias ${\cal E}^{(b)}_{s}$ adiabatically. Then, the $\pi$-light pulses are applied to resonantly couple the two-photon $5S_{1/2}(m=\frac{1}{2})-5P_{3/2}(m=\frac{3}{2})-22D_{5/2}(m=\frac{5}{2})$ transition. Afterwards, ${\cal E}_{s}$ is increased back to ${\cal E}^{(q)}_{s}$ adiabatically. (3) ${\cal E}_{s}$ goes up adiabtically to ${\cal E}^{(c)}_{s}$, where the one-photon avoided crossing occurs. After passing ${\cal E}^{(c)}_{s}$, ${\cal E}_{s}$ raises to ${\cal E}^{(a)}_{s}$ nonadiabatically. ${\cal E}_{s}$ stays at ${\cal E}^{(a)}_{s}$ for a $\frac{\pi}{2}$-pulse time length of $\pi/(2\Omega)$ and then goes back to ${\cal E}^{(q)}_{s}$ adiabatically. (4) Repeat the second step. (5) Repeat the first step.

During the qubit inversion (1), the atomic qubit switches its state while the photonic qubit remains unchanged. The light pulses in step (2) bring $|\uparrow,0\rangle$ and $|\uparrow,1\rangle$ components out of the restricted two-qubit space so that they avoid being affected by the vacuum-Rabi oscillation in the following step. Although ${\cal E}_{s}$ staying at ${\cal E}^{(a)}_{s}$ for a $\pi$-pulse time length in step (3) causes the state flipping of atom, the atom-resonator-interaction-induced anticrossing at ${\cal E}_{s}={\cal E}^{(c)}_{s}$ enables the atom to remain in $|\downarrow\rangle$ after step (3). As a result, the photonic qubit flips between $|0\rangle$ and $|1\rangle$ with the atom in $|\downarrow\rangle$. In step (4), the atom in the ground state is brought back to $|\uparrow\rangle$ without changing the photon state. Finally, to obtain the truth table of CNOT operation, the atomic qubit should switch its state again while the photonic qubit remains unchanged in step (5). The logic-gate-operation duration can be well smaller than the dephasing times of qubits.

\textit{Conclusion.} In summary, we have investigated a hybrid system composed of a Rydberg atom interacting with a SC microwave LC oscillator. Varying the external electrostatic field not only controls the qubit transition of atom but also shifts the atomic-qubit transition resonantly to the resonator, resulting in an atom-photon interaction stronger than or almost equal to that of atom-cavity, SC qubit-resonator, and atom-waveguide systems performed in recent experiments and an extremely large cooperativity. In the dressed-state picture, a pair of $|\uparrow,N-1\rangle$ and $|\downarrow,N\rangle$ can form an artificial two-state particle, leading to the SC qubit without JJs. Moreover, the universal two-qubit logic operations can be performed on this hybrid system by means of sweeping the electrostatic field and standard spectroscopic techniques.

In our physical specification, both Rabi frequency $\Omega$ and atom-resonator coupling strength $g$ are of the order of 1 GHz. All gate operations can be accomplished within a time scale much shorter than any decoherence times of LC resonator and atomic Rydberg states. Careful designing the LC resonator and choosing higher Rydberg states will further reduce $\omega_{0}$ and enhance $g$, leading to a hybrid system in the ultrastrong-coupling regime, $g\sim\omega_{0}$ or even $g>\omega_{0}$. In this case, the effects of counter-rotating terms of the atom-cavity interaction also play a major role, giving rise to more interesting physics. Moreover, the large SC resonator-atom interaction will enable the strong coupling between neutral atoms and usual SC qubits via the SC resonator.

\begin{acknowledgments}
This research is supported by the National Research Foundation Singapore under its Competitive Research Programme (CRP Award No. NRF-CRP12-2013-03) and the Centre for Quantum Technologies, Singapore.
\end{acknowledgments}

\end{document}